# Considerations on the use of financial ratios in the study of family businesses

Geòrgia Escaramís

Universitat de Girona, Research Group on Statistics, Econometrics and Health (GRECS)

Faculty of Business and Economics, Campus Montilivi, 17071 Girona

Telephone number: +34 972 41 87 28

Email address: georgia.escaramis@udg.edu

Anna Arbussà

Universitat de Girona

Faculty of Business and Economics, Campus Montilivi, 17071 Girona

Telephone number: +34 681 31 22 40

Email address: anna.arbussa@udg.edu

**Abstract**

Most empirical works that study the financing decisions of family businesses use financial ratios. These data present asymmetry, non-normality, non-linearity and even dependence on the results of the choice of which accounting figure goes to the numerator and denominator of the ratio. This article uses compositional data analysis (CoDa) as well as classical analysis strategies to compare the structure of balance sheet liabilities between family and non-family businesses, showing the sensitivity of the results to the methodology used. The results prove the need to use appropriate methodologies to advance the academic discipline.

# Considerations on the use of financial ratios in the study of family businesses

**Abstract**

Most empirical works that study the financing decisions of family businesses use financial ratios. These data present asymmetry, non-normality, non-linearity and even dependence on the results of the choice of which accounting figure goes to the numerator and denominator of the ratio. This article uses compositional data analysis (CoDa) as well as classical analysis strategies to compare the structure of balance sheet liabilities between family and non-family businesses, showing the sensitivity of the results to the methodology used. The results prove the need to use appropriate methodologies to advance the academic discipline.

**Introduction**

As highlighted in the review by Michiels & Molly (2017), financial decisions in family firms exhibit unique characteristics that set them apart from non-family firms. Notably, the emphasis on retaining control and influence, along with other non-financial values linked to family affective needs, can significantly impact the financial decisions of owners/managers (Bauxali-Soler et al., 2021; Belda-Ruiz et al., 2022; Berrone et al., 2012; Blanco-Mazagatos et al., 2024; Block et al., 2024; Camisón et al. 2022; Gómez-Mejía et al., 2007; 2011; Molly et al., 2019). Understanding the financial decisions of family firms is important because they are linked to key strategic decisions (Ginesti et al., 2023; Gómez-Mejía et al., 2018).

As Álvarez-Díaz et al. (2023) point out, financial decisions continue to be one of the most relevant areas of research in family businesses, and studies to date have provided inconclusive results on many aspects of these decisions (see, for instance, a meta-analysis by Hansen & Block, 2021; Bergmann, 2024, Comino-Jurado et al., 2021a; 2021b; Díaz-Díaz et al., 2023;

Diéguez-Soto & López-Delgado, 2019; Jansen et al., 2023). Michiels & Molly (2017) suggested further research on context-specific elements such as family business heterogeneity and country-specific factors. An example of this would be the interesting research that has more recently been produced on family involvement in the ownership and management of the firm (Álvarez-Díez et al., 2023; Belda-Ruiz et al., 2022; Blanco-Mazagatos, 2024; Hansen & Block, 2021; Jansen et al., 2023).

This article states that an additional reason for the inconclusive results obtained in the literature may be related to methodological issues. In their review of empirical studies on the accounting performance of listed family firms versus non-family firms, Heino et al. (2019) point out the methodological deficiencies of many of the studies reviewed. Moreover, in the review by Michiels & Molly (2017), most articles on financing decisions in family businesses applied regression techniques, with data coming mainly from public databases and with more than 40% of the articles discussing debt decisions, such as, for example, leverage or debt maturity, mainly using traditional financial ratios, a.k.a. accounting ratios. This article will focus on methodological deficiencies that may arise with the use of traditional financial ratios.

While traditional accounting ratios are useful for assessing the financial structure (a.k.a. capital structure) of individual companies, including family firms, their reliability becomes questionable when these ratios are used as dependent variables in statistical studies that aim to evaluate the broader economic and financial structure of family firms as a group. In fact, when used as variables in statistical evaluations, traditional financial ratios—such as indebtedness, liquidity, or debt maturity—are known to present serious statistical and practical issues that can distort statistical inference. These issues have been established for some years and include asymmetry (Faello, 2015; Frecka & Hopwood, 1983; Linares-Mustarós et al., 2022), non-linearity (Carreras-Simó & Coenders, 2021; Cowen & Hoffer, 1982), severe non-normality (Deakin, 1976; Iotti et al., 2024; So, 1987), extreme outliers (Demiraj et al., 2024; Lev and

Sunder, 1979) and even a dependence of results on the arbitrary choice of numerator and denominator in each ratio (Frecka & Hopwood, 1983; Linares-Mustarós et al., 2022).

Table 1 includes a list of recently published articles that study financing decisions in family firms and use traditional financial ratios as response variables. The fourth column of the table shows the methodology employed and the fifth column, the ratios used as dependent variables. Next to it the table indicates financial ratios used as independent, moderating or control variables. Finally, the last column indicates the variables that have been transformed (usually, age and size are transformed with natural logarithms), recognizing this need when dealing with variables that do not present normal distributions or have outliers. However, similar transformations are not applied to financial ratios, even though these ratios do not meet the desirable characteristics to be used as variables in various statistical techniques, without having been previously transformed.

"Insert Table 1 about here"

An alternative approach to the use of traditional financial ratios involves the use of new accounting ratios derived from the compositional data analysis methodology, commonly referred to as Compositional Data (CoDa). The validity of this methodology has been widely validated in previous research (Aitchison, 1986; Pawlowsky-Glahn et al., 2015; Coenders et al., 2023). Although CoDa originated in the fields of geology and chemistry at the end of the last century, primarily to assess the relative importance of components in chemical analyses, its applications have since expanded to various scientific domains, including economics, business, and other social sciences. The use of CoDa to analyze accounting ratios is a more recent development (Arimany-Serrat et al., 2022; 2023; Arimany-Serrat & Sgorla, 2024; Carreras-Simó & Coenders, 2020; 2021; Creixans et al., 2019; Dao et al., 2024; Jofre-Campuzano & Coenders, 2022; Linares-Mustarós et al., 2018; 2022; Molas-Colomer at al., 2024; Saus-Sala et al., 2024).

This study compares the analysis of the financial structure of family versus non-family firms using traditional and compositional ratios. It aims to demonstrate significant differences in the results, highlighting the need to revise conventional methodologies employed in the study of family firms with statistical methods. In this way, methodological improvements in family firm research would be on par with those in other areas of management research (Carreras-Simó and Coenders, 2021; Ferrer-Rosell et al., 2021; Grifoll et al., 2019; Joueid & Coenders, 2018; Morais et al., 2018; Sieber, 2024, Wang et al., 2019).

The remainder of this article is structured in three main sections. First, significant problems arising from the use of traditional ratios in statistical studies of family firms are discussed and it is demonstrated how accounting ratios based on the CoDa methodology (hereafter, compositional accounting ratios) address these problems. The next section uses data extracted from the Survey on Business Strategies (ESEE) conducted by the SEPI Foundation, an annual panel survey focused on manufacturing firms in Spain. This section compares the analysis of family and non-family firms using both traditional and compositional ratios, highlighting the substantial differences in the results to underline the need to review the conventional methodologies used in accounting studies in this context. The following section discusses the results and concludes.

**Methodology**

The loss of symmetry that occurs when defining traditional accounting ratios causes significant distortions in the evaluation and identification of factors that influence the financial health of firms. However, as Linares-Mustarós et al. (2022) point out, it has not received the attention it deserves. The fact that ratios have values over the interval $(0, +\infty)$ can only cause asymmetric distributions, since the values of the ratio when the denominator exceeds the numerator are in the interval $(0, 1)$, and the values where the numerator exceeds the denominator are in the interval (1, +∞). Moreover, the authors describe in detail the consequences, for example, of reversing the coefficients of a ratio. If we have two balance sheet values, $x_1$ and $x_2$, which we assume to be strictly positive, as for example $x_1$= long-term liabilities and $x_2$ = short-term liabilities, the ratios $x_1/x_2$ and $x_2/x_1$ should provide equivalent information on debt maturity. It is crucial to exercise caution when analyzing them with statistical purposes. To illustrate this problem, we present a toy example that highlights the potential distortions in analyzing balance sheet data of firms. In this example, we created a dataset comprising 10 fictitious companies, each with two positive balance sheet values: $x_1$ and $x_2$. These values were generated randomly, and from them, two traditional ratios were calculated: ratio A $= \frac{x_1}{x_2}$ and the permuted version ratio B $= \frac{x_2}{x_1}$ (see Figure 1A).

"Insert Figure 1 about here"

Figure 1B displays boxplots to compare the two ratios across the 10 companies. As shown in the graph, Company 4 appears to be an outlier when considering Ratio A. In contrast, for the reverse ratio, Company 3 appears as the outlier. However, both companies are almost equally distanced from the rest when examined closely. Following this analysis, it is also evident that, as expected given the arithmetic nature of financial ratios, the dispersion among lower values is substantially smaller than that among higher values. This variation further emphasizes the need for caution when interpreting financial ratios, as it can skew perceptions of company performance depending on the ratio structure and chosen denominator. Such discrepancies highlight the potential for confusion and misinterpretation when traditional accounting ratios are applied interchangeably, underscoring the need for a more standardized approach such as CoDa.

CoDa is linked to the study of relative magnitudes expressed in ratio form and constitutes a robust alternative to classical financial statement analysis. This is particularly relevant in the context of family firms. Given the unique financial structures often found in family firms, applying CoDa and the corresponding compositional accounting ratios to their financial analysis could reveal insights that traditional methods might overlook, providing a clearer understanding of their financial health and performance.

In the remainder of this section we show how the compositional accounting ratios are constructed. To represent financial ratios using a composition of an enterprise's balance sheet values, the first step is to define the specific financial aspect we aim to measure. Once this is established, we can construct ratios, accordingly, following a similar approach to traditional methods. Suppose our objective is to focus on the aggregated categories in the balance sheet. In this case, a composition with $D = 4$ components comes into play: $x_1$ = long-term liabilities; $x_2$ = short-term liabilities; $x_3$ = fixed assets; $x_4$ = current assets.

Based on this framework, we can define three traditional financial ratios. The first is an indebtedness ratio. This ratio divides long-term plus short-term liabilities by fixed plus current assets: $r_1 = \frac{x_1+x_2}{x_3+x_4}$. The second is a debt maturity ratio: $r_2 = \frac{x_1}{x_2}$. The third is an asset tangibility ratio, $r_3 = \frac{x_3}{x_4}$. The first ratio is often permuted becoming a measure of solvency of the firm: $r_{1p} = \frac{x_3+x_4}{x_1+x_2}$.

The CoDa methodology evaluates the relative importance of accounting figures using logarithms and geometric averages. Unlike the arithmetic average, which is incompatible with ratio-based analyses, geometric averages are fundamental to CoDa, as they allow for meaningful comparisons between magnitudes in relative terms. Arimany-Serrat & Sgorla (2024) illustrate this concept through a simple example; the ratio between 81 and 27 (81/27 = 3) is equivalent to the ratio between 27 and 9 (27/9 = 3), establishing 27 as the central value between 9 and 81 in relative terms. Here, the geometric average of 9, 27, and 81 is 27, calculated as the cube root of the product $9 \times 27 \times 81$. Conversely, the arithmetic average of these values is 39, which skews toward the largest absolute value, 81.

Moreover, in our context, when comparing financial ratios between family and non-family firms, the geometric average proves especially useful for maintaining consistency in aggregated financial results across both categories. In comparing financial ratios for family versus non-family firms, calculating averages with the geometric average helps prevent distortions that can arise if arithmetic means are used. This is because the geometric average preserves the proportional relationship between individual company ratios, enabling a more accurate and representative comparison across both groups. For instance, when calculating the geometric average of financial ratios for family and non-family firms, the result better reflects the aggregate behavior of each group in this metric. Conversely, with the arithmetic mean, companies with extremely high values (usually in the non-family group) could distort the results, making the comparison less representative of the actual financial profile of each group.

The usual CoDa approach is to use existing standard statistical methods on transformed data with logarithms of ratios of geometric averages. The transformed data is based on the principle of that for each composition $D - 1$ log-ratios (in our example $D = 4$) are constructed as balance coordinates. These coordinates are formed from a sequential binary partition (SBP) of components as shown in Figure 2 (Egozcue & Pawlowsky-Glahn, 2005).

"Insert Figure 2 about here"

To create the first coordinate, the complete composition $x = (x_1, x_2, \ldots, x_D)$ is divided into two groups of components: one for the numerator and the other for the denominator of the log-ratio. In the next step, one of these two groups is further divided to create the second coordinate, and the process continues iteratively. The coordinates are scaled log-ratios of the geometric

averages of each group of components. In our example, three log-ratios can be defined as follows:

$$z_1 = \sqrt{\frac{4}{4}} \log\left(\frac{\sqrt{x_1 \cdot x_2}}{\sqrt{x_3 \cdot x_4}}\right)$$

$$z_2 = \sqrt{\frac{1}{2}} \log\left(\frac{x_1}{x_2}\right)$$

$$z_3 = \sqrt{\frac{1}{2}} \log\left(\frac{x_3}{x_4}\right)$$

We observe that the first compositional accounting ratio $z_1$ includes all the accounting figures to be studied and separates the liabilities (numerator) from the assets (denominator) in the same manner as the indebtedness ratio. Furthermore, multiplying by the logarithm introduces a scaling factor that does not alter the interpretation of the ratio and is used to account for the number of accounting figures being compared. Its denominator is the total number of accounting figures involved in the ratio $(2 + 2 = 4)$, while its numerator is the product of the number of accounting figures appearing in the denominator and the numerator $(2 \times 2 = 4)$. On the other hand, the compositional accounting ratios $z_2$ and $z_3$ contain the two more disaggregated accounting figures from the tree diagram in Figure 2, along with their corresponding scaling factor, and correspond to the ratios of debt maturity and asset tangibility, respectively.

This section highlights several key observations regarding the use of compositional accounting ratios. First, the selection of the tree shown in Figure 2 is left to the researcher's discretion. In our case, we chose the tree to derive three compositional ratios relevant to the concepts of interest or to mirror known traditional ratios. Additionally, it is important to note that swapping the numerator and denominator of a compositional accounting ratio only affects its sign. This property ensures that the same outlier values remain consistent across both configurations. Furthermore, this feature guarantees that relationships with non-accounting indicators—such as mean differences, correlations, and regression coefficients—retain the same magnitude, albeit with opposite signs. This property also allows the ratios to generate values spanning the entire range from negative to positive infinity, closely resembling a normal probability distribution.

**Results**

The data:

The data used in this section has been extracted from the Survey on Business Strategies (ESEE) conducted by the SEPI Foundation (https://www.fundacionsepi.es/). The ESEE is an annual panel survey focusing on manufacturing firms situated in Spain. The survey's geographical scope encompasses the entirety of the Spanish economy, and it employs yearly variables. It targets manufacturing companies with 10 or more employees. The ESEE aims to gather information about firms' strategies, specifically the decisions they make regarding various competition variables, from prices to research and development (R&D) expenditures. It also includes the firms' balance sheet together with their profit and loss statements. We obtained a total of 1,031 companies, of which 521 are non-family firms and 381 are family-firms.

The variables selected for the analysis include family or non-family firm (firms in which owners and / or other family members work at the firm), technology intensity (3 categories according to NACE), product innovation (the company obtained product innovations), firm size (log of the number of employees) and financial statements of the firm. These variables were collected annually, covering the period from 2007 to 2018, resulting in panel data for the analysis.

The components analyzed to assess the financial structure of family firms are derived from the balance sheet and consist of $D = 3$ positive non-overlapping account categories: short-term liabilities ($x_1$), long-term liabilities ($x_2$) and shareholders' equity ($x_3$). These accounting figures make it possible to compute some useful indicators of financial structure, like the traditional ratio of short-term liabilities over long-term capital, from now on, financial stability ratio, $r_1 = \frac{x_1}{x_2+x_3}$, and the long-term indebtedness ratio, $r_2 = \frac{x_2}{x_3}$. High values for $r_1$ would identify firms with low financial stability. Conversely, in the permuted version of the ratio ($r_{1p} = \frac{x_2+x_3}{x_1}$), high values signify firms with high financial stability. For the case of $r_2$, high values indicate firms with high long-term indebtedness. However, on the permuted version ($r_{2p} = \frac{x_3}{x_2}$), high values correspond to firms with low long-term indebtedness.

As outlined in the methodology, the compositional accounting ratios are constructed using balance coordinates derived through the SBP process. The SBP is illustrated in Figure 3, where the resulting coordinates are defined as follows to mirror the traditional ratios:

$$z_1 = \sqrt{\frac{2}{3}} \log\left(\frac{x_1}{\sqrt{x_2 \cdot x_3}}\right)$$

$$z_2 = \sqrt{\frac{1}{2}} \log\left(\frac{x_2}{x_3}\right)$$

"Insert Figure 3 about here"

Figure 4 presents the distribution of the analysed ratios, separated by family and non-family firms. Outliers are displayed as individual points. Panels A) and B) show the first accounting ratio ($r_1$) and its permuted version ($r_{1p}$), respectively, while panels C) and D) display the second accounting measure ($r_2$) and ($r_{2p}$).

The distribution pattern of the traditional accounting ratios is noticeably asymmetric. In all cases, the asymmetry is positive, regardless of whether the original or permuted version of the accounting measure is used. This leads to contradictory conclusions in the interpretation. For example, panel A) suggests that firms with lower financial stability are more dispersed than those with higher stability (i.e., the highest ratio values are more spread out than the lowest). In contrast, panel B) reverses this interpretation, showing that more stable firms appear to be the most dispersed. A similar inconsistency is observed in panels C) and D), which represent the long-term solvency ratio.

In contrast, panels E), F), G), and H) display the compositional accounting ratios. These panels exhibit a clear symmetric distribution, and the interpretation of the data remains consistent, regardless of which figure is in the numerator or denominator. The permuted boxplots in panels G) and H) are the original boxplots in panels E) and F) upside down. Both traditional and compositional ratios have outliers, but in traditional ratios there are much more of them, they are much more extreme and tend to dominate the distribution of the ratio.

"Insert Figure 4 about here"

To analyze the differences in financial structure, measured through accounting ratios, between family and non-family firms, mixed-effects models were employed. This approach controls not only the confounding variables, such as technology intensity, product innovation, and firm size, but also the panel structure of the data collected annually from 2007 to 2018. The mixed-effects model allows for capturing both fixed effects—reflecting the influence of explanatory

variables—and random effects—accounting for unobserved heterogeneity across firms and the correlation structure within repeated observations over time.

The models:

Separate models were specified to analyse the relationships between the explanatory variables and different dependent variables representing financial stability and long-term indebtedness. For each dependent variable, both the traditional and compositional financial ratios were used. The models were expressed as follows:

$$y_{it} = \beta_0 + \beta_1 Family_{it} + \beta_2 MildTechIntens_{it} + \beta_3 HighTechIntens_{it} + \beta_4 Innovation_{it} + \beta_5 FirmSize_{it} + \sum_{j=2008}^{2018} \beta_j YearDummy_{it} + u_i + \varepsilon_{it}$$

Where:

- $y_{it}$: The dependent variable for firm $i$ at time $t$, which alternates between the traditional and compositional versions of financial stability and long-term indebtedness.
- $Family_{it}$: A binary variable indicating whether firm $i$ is a family firm (1) or not (0).
- $MildTechIntens_{it}$ and $HighTechIntens_{it}$: Control variables representing the firm's level of technological intensity.
- $Innovation_{it}$: A control variable indicating whether firm $i$ offers innovative products (1) or not (0).
- $FirmSize_{it}$: A control variable for firm size, constructed as the log of the number of employees in the firm.
- $YearDummy_{it}$: A set of dummy variables representing the year of observation for firm $i$ at time $t$, with 2007 as the baseline year (0).

- $u_i$: The random effect at the firm level, accounting for unobserved heterogeneity between firms and allowing for the correlation structure of firm observations over time.
- $\varepsilon_{it}$: The residual error term.

Table 2 presents the parameter estimates of the models used in the analysis. Regarding the models based on traditional ratios, the results are inconsistent depending on whether the original or permuted version of the ratio is used. For instance, in the case of long-term indebtedness ($r_2$), the results suggest that larger firms appear to be more indebted in the long term, as indicated by the positive coefficient ($\beta=0.177$) and a significant p-value at $\alpha=10\%$ ($p=0.077$). However, when using the permuted version of the ratio, the coefficient remains positive ($\beta=451.31$), but in this context, it implies that larger firms are less indebted, although the effect is not statistically significant ($p=0.236$). Sign inconsistencies also appear for mid and high technical intensities.

Table 2 also summarizes the results for the models using compositional ratios. In terms of financial stability ($z_1$), our findings suggest that family firms exhibit greater financial stability ($p=0.068$). Conversely, firms with higher technological intensity tend to show reduced financial stability ($p=0.041$; $p=0.001$), and so do larger firms ($p=0.029$). For long-term indebtedness ($z_2$), the analysis does not reveal significant differences between family and non-family firms ($p=0.215$). However, firms with higher technological intensity are associated with lower indebtedness ($p=0.010$; $p=0.046$) an so do innovative firms ($p=0.027$), while larger firms tend to exhibit higher indebtedness ($p=0.080$).

The permuted versions of the compositional ratios are not displayed in Table 2, as their parameter estimates are identical to those of the original ratios but with reversed signs. Consequently, no inconsistencies arise when switching the numerator and denominator in compositional ratios.

The results obtained with compositional ratios fundamentally differ from those obtained with traditional ratios. Only one significant effect emerges for $r_2$ and none for $r_1$, $r_{1p}$, and $r_{2p}$.

Figure 5 illustrates the residual behaviour of the models. For the traditional ratios ($r_1, r_{1p}, r_2$ and $r_{2p}$), a clear pattern of heteroscedasticity and extreme outliers are observed. In contrast, the residuals for the compositional models are randomly distributed around the zero line, which aligns with the assumptions of the model. The differences observed between traditional and compositional ratios in Table 2 can thus be attributed to the former not fulfilling the model's assumptions, which leads to disregarding them.

"Insert Table2 about here"

"Insert Figure 5 about here"

**Discussion and conclusion**

The results of the previous section show the limitations of using traditional financial ratios as dependent variables in statistical studies intended to evaluate groups of firms. For their use to be appropriate, the requirements of symmetry, linearity, absence of outliers, and normality must be met, which has shown not to be the case throughout this work. In addition, inconsistencies have been revealed between the results that arise from their use when reversing the positions between numerator and denominator.

The use of the CoDa methodology is suggested to transform traditional financial ratios so that the new ratios do meet the assumptions required for their use in statistical studies. Mixed effects models have been used to analyze the differences in the financial structure, measured through accounting ratios, between family and non-family companies. The results obtained differ depending on whether traditional or compositional financial ratios are used. The clear pattern of heteroscedasticity and outliers of the model's residuals in the first case is indicative of its poor validity.

Although these limitations have been established in the use of traditional financial ratios in different areas of study (Arimany-Serrat et al., 2022; Carreras-Simó & Coenders, 2021; Creixans-Tenas et al., 2019; Dao et al., 2024; Jofre-Campuzano & Coenders, 2022; Linares-Mustarós et al., 2018; 2022), their use continues to be relevant in the field of family business, as may be seen in Table 1. This may be due to their acceptance in prestigious journals. A possible, albeit speculative, explanation for this may be that, as researchers, we tend to replicate some aspects of the works we consult. An example of this could be the logarithmic transformation of the variables size and age to resolve their asymmetry, in almost all the articles in the sample selected in Table 1 and, instead, not considering the asymmetries of the ratios.

Methodological improvements can contribute to clarifying the diversity of results that has been observed to date in the analysis of the financial structure of family businesses, either when comparing them with non-family businesses, or by considering family businesses as a heterogeneous group. It should be noted that, in this sense, in the literature in the field of family business, very interesting works have been published in recent years that help to understand this diversity that occurs among family businesses. It is fair to point out that most of these studies use a much larger number of variables than the one used in the analysis in our article and, in doing so, manage to introduce more nuances into the analysis of family business behavior. This work does not aim to compare with this literature in this respect. However, this article contributes to the literature by highlighting the methodological problems that arise with the use of traditional accounting ratios in statistical analysis. The correctness of the methodology is crucial to establish the veracity and reliability of the results.

**Figure 1.** Traditional financial ratios calculated for 10 fictitious companies based on two positive balance sheet values $x_1$ and $x_2$. Panel A: raw data and Panel B: boxplot illustrating the distribution of the two ratios for the ten fictitious companies.

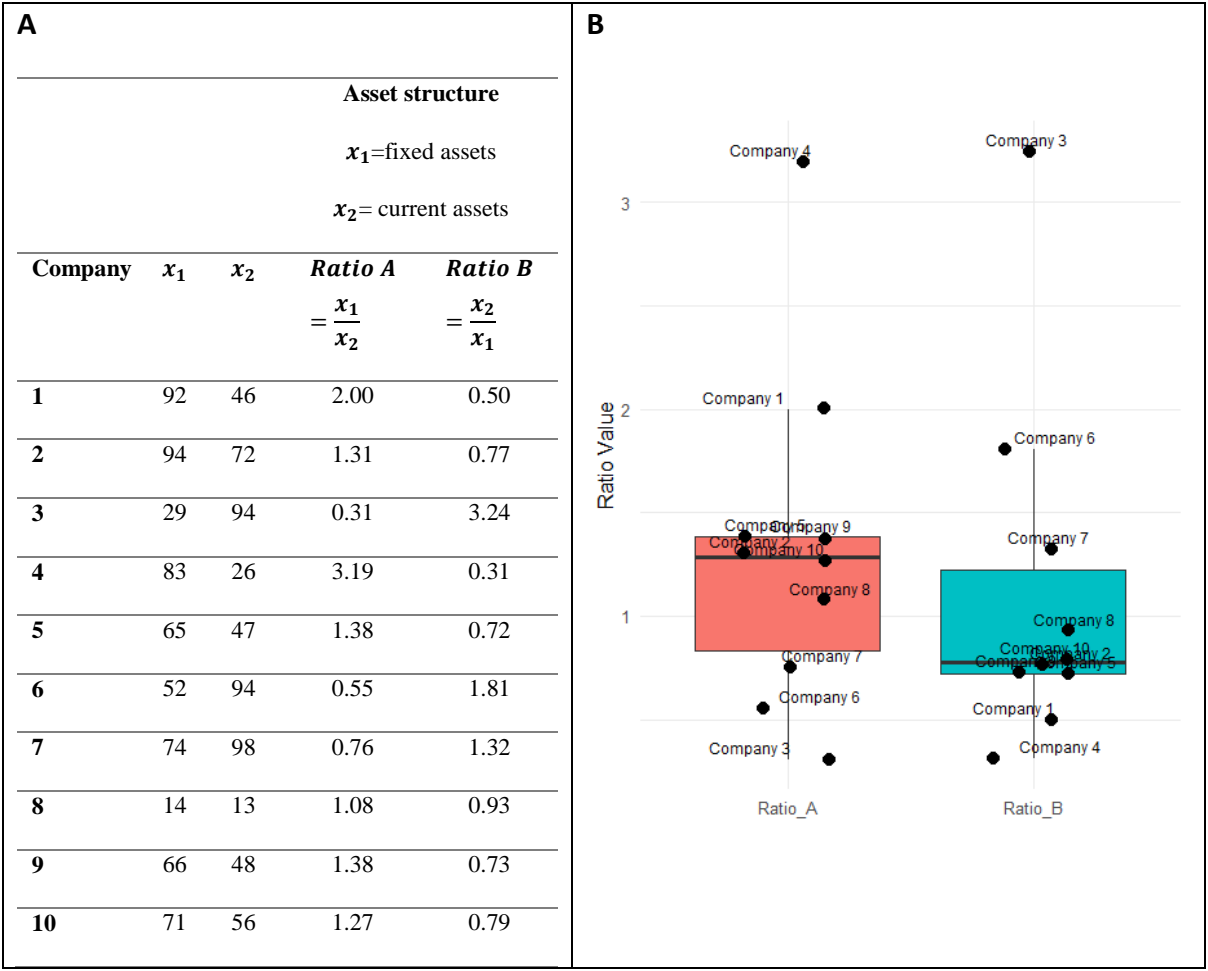

| Company | $x_1$ | $x_2$ | Ratio A $= \frac{x_1}{x_2}$ | Ratio B $= \frac{x_2}{x_1}$ |
|---|---|---|---|---|
| | | | Asset structure $x_1$=fixed assets $x_2$= current assets | |
| 1 | 92 | 46 | 2.00 | 0.50 |
| 2 | 94 | 72 | 1.31 | 0.77 |
| 3 | 29 | 94 | 0.31 | 3.24 |
| 4 | 83 | 26 | 3.19 | 0.31 |
| 5 | 65 | 47 | 1.38 | 0.72 |
| 6 | 52 | 94 | 0.55 | 1.81 |
| 7 | 74 | 98 | 0.76 | 1.32 |
| 8 | 14 | 13 | 1.08 | 0.93 |
| 9 | 66 | 48 | 1.38 | 0.73 |
| 10 | 71 | 56 | 1.27 | 0.79 |

**Figure 2.** Tree structure obtained from a Sequential Binary Partition (SBP) of components. The components are: $(x_1)$ long-term liabilities, $(x_2)$ short-term liabilities, $(x_3)$ fixed assets, and $(x_4)$ current assets.

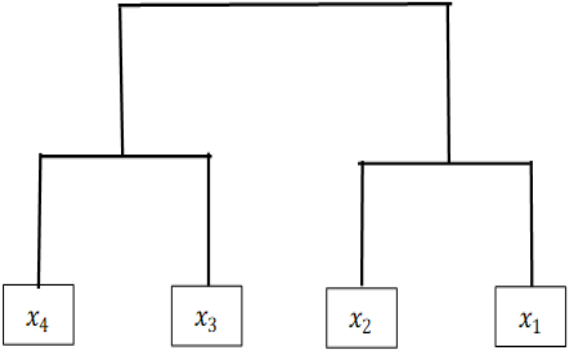

**Figure 3.** Tree structure obtained from a Sequential Binary Partition (SBP) of components. The components are: ($x_1$) short-term liabilities, ($x_2$) long-term liabilities, and ($x_3$) shareholders' equity.

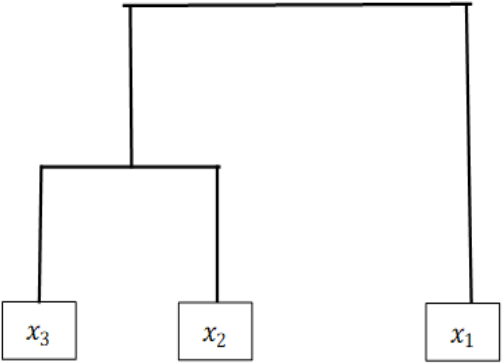

**Figure 4.** Boxplots illustrating the distribution of the dependent variables, including traditional financial ratios and their compositional versions. The plots compare the original variables with their permuted numerator-denominator versions. Panel (A) represents the financial stability traditional ratio, (B) the financial stability with permuted numerator and denominator, (C) the long-term indebtedness traditional ratio, (D) the long-term indebtedness with permuted numerator and denominator, and (E–H) correspond to panels (A–D) with the compositional ratios.

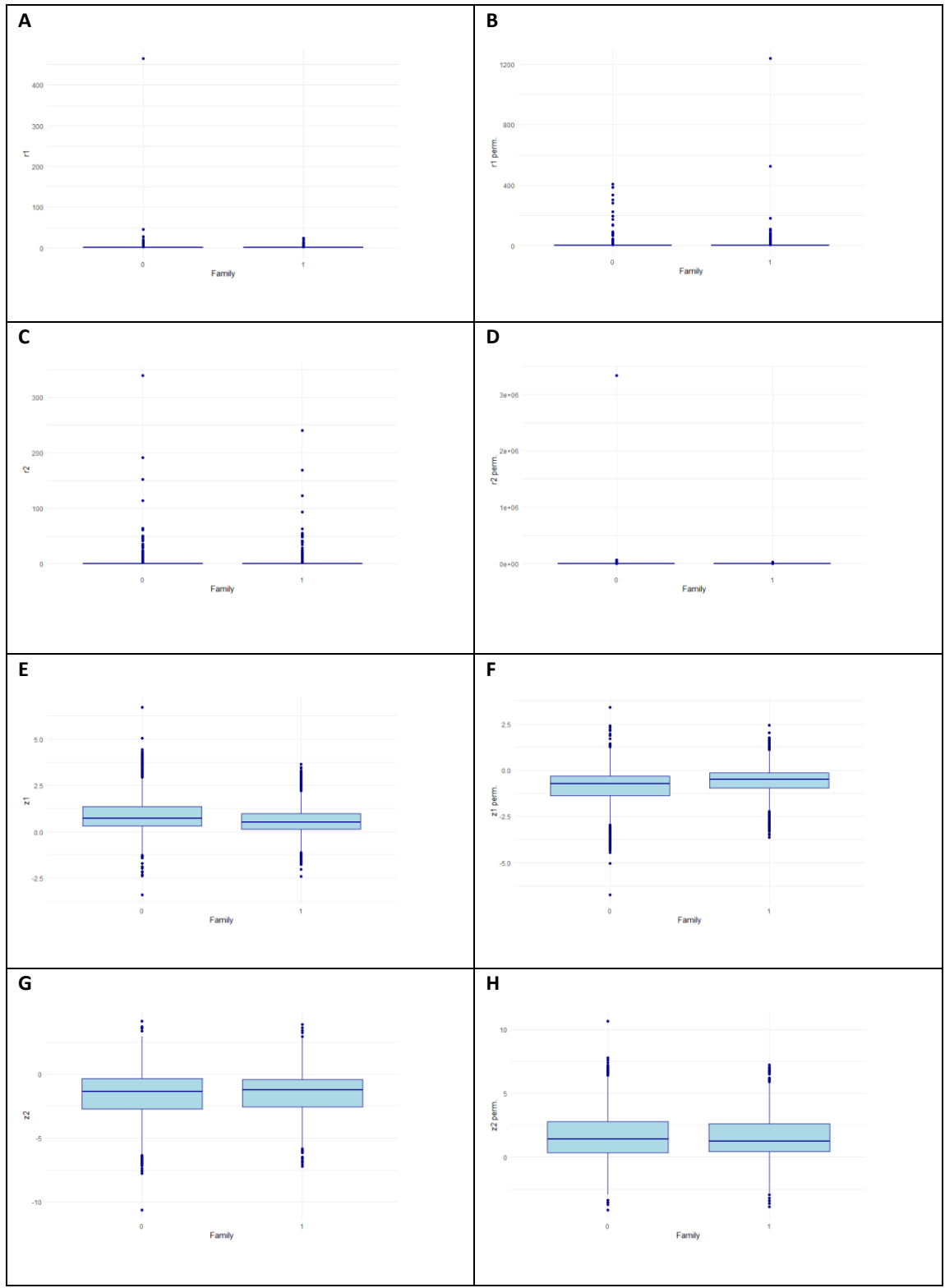

**Figure 5.** Predicted values versus residuals plots for each analyzed financial ratio. Panel (A) represents the financial stability traditional ratio, (B) the long-term indebtedness traditional ratio, (C) the financial stability with permuted numerator-denominator version, (D) the long-term indebtedness with permuted numerator-denominator version, (E) the compositional financial stability ratio, and (F) the compositional long-term indebtedness ratio.

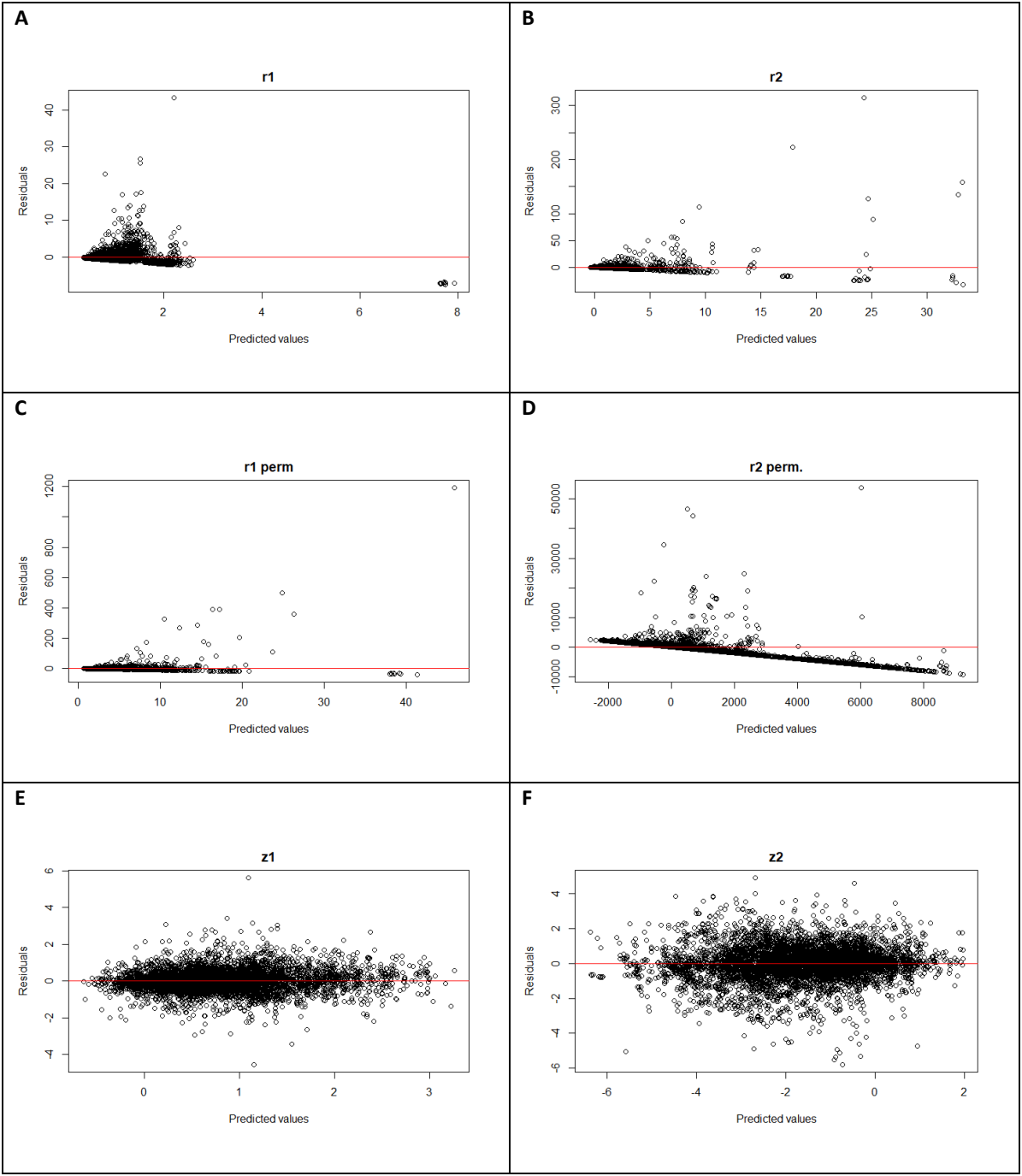

Table 1. Selected articles from the field of family business that use financial ratios as dependent variables

| Year | Authors | Citations (Google scholar at 20/01/2025) | Methodology | Response variables | Independent, moderating and control variables include | Transformed variables |
|---|---|---|---|---|---|---|
| 2024 | Al-Haddad et al. | 10 | Linear regression models | ROA, Tobin's Q | Various financial ratios | Size |
| 2018 | Bacci et al. | 43 | Linear regression model for panel data | Non-equity liabilities over total assets | Various financial ratios | Size |
| 2021 | Baixauli-Soler et al. | 73 | Linear regression models | Debt over total assets | Various financial ratios | Size and Age |
| 2024 | Blanco-Mazagatos et al. | 4 | Linear regression models | Financial debt over assets | Various financial ratios | Size and Age |
| 2024 | Block et al. | 4 | Linear regression models | Debt ratio and long-term debt ratio | Various financial ratios | Size and Age |
| 2022 | Camisón et al. | 29 | Linear regression model for panel data | Leverage (long-term plus short-term financial debt over total assets) | Various financial ratios | Size and Age |
| 2021a | Comino-Jurado et al. | 10 | Partial least square models | Indebtedness (total liabilities over total assets) | Family ownership ratio | Size and Age |
| 2021b | Comino-Jurado et al. | 42 | Partial least square models | Indebtedness (debt over total assets) | Various financial ratios | Size and Age |
| 2023 | Díaz-Díaz et al. | 4 | Linear regression model for panel data | Bank debt over total assets | Various financial ratios | Size and Age |
| 2019 | Diéguez-Soto & López-Delgado | 42 | Linear regression models | Leverage (total debt over net assets and total liabilities over total assets) | Various financial ratios | Size and Age |
| 2023 | Ginesti et al. | 8 | Linear regression models | Debt maturity (long-term debt over total debt) | Various financial ratios | Size |
| 2020 | López-Delgado & Diéguez-Soto | 41 | Linear regression model for panel data | Indebtedness (total debt over total assets and total liabilities over total assets) | ROA | Size and Age |
| 2019 | Molly et al. | 122 | Linear regression models | Various debt ratios (total debt over total assets, long-term debt over total assets, short-term debt over total assets, interest-bearing debt over total assets and non-interest bearing rate over total assets) | Various financial ratios | Size |
| 2024 | Muñoz-Bullón et al. | 5 | Linear regression models | Leverage (long-term plus short-term financial debt over total assets) | Various financial ratios | Size and Age |
| 2022 | Serrasqueiro et al. | 9 | Linear regression models | Differences in leverage ratios (target ratios minus short-term debt over total assets and long-term debt over total assets) | Various financial ratios | Size and Age |

**Table 2.** Coefficient estimates, standard errors (SE), and p-values from the mixed-effects model for the variables *Family*, *TechInt (Mid)*, *TechInt (High)*, *Innovation*, and *Size (log-persons)*. The estimates reflect the effects of these predictors on the dependent variables, which include the traditional financial stability ratio ($r_1$), the permuted numerator-denominator version ($r_{1p}$), the compositional financial stability ratio ($z_1$), the traditional long-term indebtedness ratio ($r_2$), the permuted numerator-denominator version ($r_{2p}$), and the compositional long-term indebtedness ratio ($z_2$).

|  | $z_1$ | | | $r_1$ | | | $r_{1p}$ | | |
|---|---|---|---|---|---|---|---|---|---|
|  | Coefficients | SE | p-value | Coefficients | SE | p-value | Coefficients | SE | p-value |
| **Family** | -0.049 | 0.027 | 0.068 | -0.263 | 0.166 | 0.114 | 0.061 | 0.621 | 0.921 |
| **TechInt. (Mid)** | 0.109 | 0.053 | 0.041 | -0.057 | 0.181 | 0.754 | 1.062 | 0.695 | 0.126 |
| **Tech Intens (High)** | 0.191 | 0.056 | 0.001 | -0.022 | 0.188 | 0.910 | 0.389 | 0.721 | 0.590 |
| **Innovation** | 0.040 | 0.024 | 0.106 | -0.257 | 0.192 | 0.182 | 0.644 | 0.702 | 0.359 |
| **Size(log-persons)** | 0.034 | 0.015 | 0.029 | 0.032 | 0.058 | 0.573 | -0.284 | 0.220 | 0.197 |
|  | $z_2$ | | | $r_2$ | | | $r_{2p}$ | | |
|  | Coefficients | SE | p-value | Coefficients | SE | p-value | Coefficients | SE | p-value |
| **Family** | 0.061 | 0.049 | 0.215 | 0.009 | 0.255 | 0.973 | -682.68 | 1108.22 | 0.538 |
| **TechInt. (Mid)** | -0.294 | 0.114 | 0.010 | 0.072 | 0.329 | 0.827 | 1485.43 | 1181.81 | 0.210 |
| **Tech Intens (High)** | -0.241 | 0.121 | 0.046 | -0.253 | 0.341 | 0.458 | -435.62 | 1225.95 | 0.722 |
| **Innovation** | -0.096 | 0.043 | 0.027 | -0.177 | 0.264 | 0.503 | 713.026 | 1306.34 | 0.585 |
| **Size(log-persons)** | 0.080 | 0.031 | 0.011 | 0.177 | 0.100 | 0.077 | 451.31 | 380.97 | 0.236 |